\documentstyle[11pt,twoside,newpasp,psfig,epsf]{article}

\newcommand{\gtap}{\mathrel{\hbox{\rlap{\lower.55ex \hbox {$\sim$}}
                   \kern-.3em \raise.4ex \hbox{$>$}}}}
\newcommand{\ltap}{\mathrel{\hbox{\rlap{\lower.55ex \hbox {$\sim$}}
                   \kern-.3em \raise.4ex \hbox{$<$}}}}

\begin{document}

\title{X-ray sources in $\omega$~Centauri and other globular clusters}

\author{Frank Verbunt}
\affil{Astronomical Institute, Utrecht University, Postbox 80.000,
   3508 TA Utrecht, The Netherlands; email verbunt@phys.uu.nl}

\begin{abstract}
X-ray bursts from bright sources in globular clusters of our galaxy show that
at least 11 (of a total 13) of these sources are neutron stars.  One of the
low-luminosity X-ray sources in $\omega$~Cen is a neutron star
accreting at a low rate. Together with the discoveries that M~15
contains two bright X-ray sources, and that 47~Tuc and NGC~6440
contain 2 and 5 low-luminosity X-ray sources with a neutron star,
this indicates that the total number of binaries with neutron stars in
globular clusters is higher than previously suspected. The discovery
of very bright X-ray sources in globular clusters near other galaxies
indicates that these clusters contain binaries with accreting black
holes, and multiple bright X-ray sources. Dozens of low-luminosity
X-ray sources have been discovered in $\omega$~Cen, NGC~6397,
NGC~6752, NGC~6440, and a hundred in 47~Tuc.  Accurate Chandra
positions of these sources combined with HST observations have lead to
unambiguous identifications with cataclysmic variables, RS CVn
binaries, and millisecond radio pulsars.
\end{abstract}


\section{Introduction}

From the first maps of the X-ray sky it was already apparent 
that globular clusters are special also in X-rays.
Of the $\sim$200 permanent or transient bright X-ray sources
in our galaxy, thirteen are now known to reside in a
globular cluster.
More sources were discovered with more sensitive instruments,
and we have just crossed the brink to a spate of discoveries
with Chandra/AXAF and with Newton/XMM.
I start this review with a brief survey of the types of X-ray
sources that may be detected in globular clusters, and 
their various formation mechanisms.
I then discuss the bright ($L_{\rm x}\gtap10^{36}$ erg\,s$^{-1}$)
cluster sources in our galaxy, in particular their preference 
for dense cores and the proof that they are accreting neutron stars,
before assessing the change in perspective due to the first Chandra
observations of globular clusters in our own and other galaxies.
Finally, I describe the dim ($L_{\rm x}\ltap10^{35}$ erg\,s$^{-1}$)
sources from a ROSAT census, and compare the Chandra observations of
$\omega$\,Cen -- described in more detail by Cool in these
proceedings -- with those of other clusters.

\section{Typology and formation mechanisms}

\begin{figure}[]
\centerline{\psfig{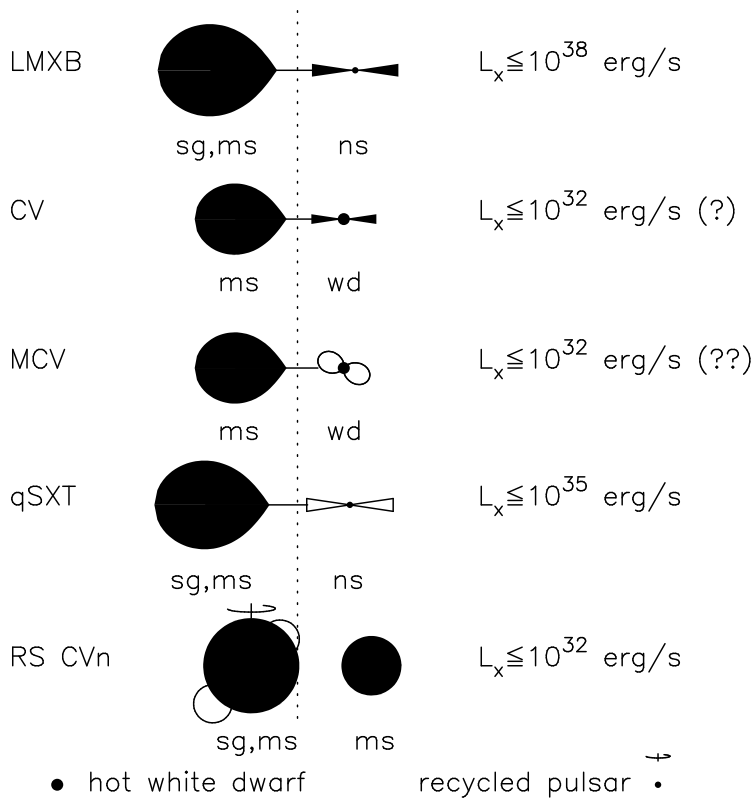}}
\caption{The various types of objects suggested
as X-ray sources in globular clusters. Binaries from top to bottom:
low-mass X-ray binaries, cataclysmic variables, magnetic cataclysmic
variables, quiescent soft X-ray transients -- in all of these
the luminosity is due to mass transfer -- and RS CVn binaries,
in which magnetic activity produces the X-rays.
Luminosies in the 0.5-2.5 keV range are indicated on the right.
Single stars that emit X-rays include hot white dwarfs, and
recycled pulsars. Recycled pulsars also occur in binaries,
with an undermassive white dwarf companion. sg, ms, ns and wd stand for
(sub)giant, main-sequence star, neutron star and white dwarf.
\label{fvfiga}}
\end{figure}

The various types of X-ray sources that one may expect to
find in a globular cluster are depicted in Fig.\,\ref{fvfiga}.
These types are selected on the basis of their luminosity,
as observed in the galactic disk where unambiguous identifications
are possible also for less accurate X-ray positions, 
with the added criterion that they occur in an old
population. This latter constraint excludes objects like
supernova remnants or T Tau stars.

The brightest sources, with ($L_{\rm x}\gtap10^{36}$ erg\,s$^{-1}$),
are neutron stars or black holes accreting from a
main-sequence or (sub)giant companion star.
Such {\em low-mass X-ray binaries} are the only option for
the bright X-ray sources.
For the dimmer sources, various types are possible.
Some neutron stars and black holes accrete at high rates only during 
outbursts that typically last a month. Between such outbursts, such
{soft X-ray transients} have much lower luminosities, down to
$10^{32}$ erg\,s$^{-1}$ for accreting neutron stars, and
to $10^{30}$ erg\,s$^{-1}$ for accreting black holes (e.g.\ Rutledge et
al.\ 2000).  
White dwarfs may also accrete matter from a companion; such
binaries are called {\em cataclysmic variables}, or {\em magnetic 
cataclysmic variables} if the white dwarf has a strong magnetic field
which disrupts (part of) the accretion disk.
The X-ray luminosities of cataclysmic variables in
the galactic disk (Verbunt et al.\ 1997)
are difficult to determine, because of uncertain distances.
In addition, most of the luminosity may be emitted at soft 
($\ltap0.5$\,keV) X-ray energies
and in the extreme ultraviolet, strongly affected by
interstellar absorption, which often is difficult to quantify.
To minimize this problem, all X-ray luminosities in this
paper are in the range 0.5--2.5 keV, unless specified otherwise.
The orbital periods of all these binaries
are on the order of hours for main-sequence donors; and
of days for (sub)giant donor stars.

In close binaries, stars can obtain or retain rapid rotation velocities
even at old age, and thus become or remain magnetically active.
Such magnetically active binaries are called {\em RS CVn} systems, 
and are X-ray emitters (Dempsey et al.\ 1993). 
In the strict definition RS CVn
systems contain a (sug)giant; I will use a wider definition,
which includes main-sequence binaries, also called BY Dra systems.

Two types of single stars complete the list of possible sources of 
X-rays: hot white dwarfs, and recycled radio pulsars.
The emission of hot white dwarfs is very soft, and will be detectable
only in globular clusters with very small interstellar absorption
(compare the work on M~67 by Belloni et al.\ 1998).
Recycled pulsars are neutron stars that once accreted matter from a binary
companion, and by this process were spun up to rapid rotation
($\ltap 0.1$\,s); for this reason they are often referred to as
millisecond pulsars. When the accretion ended, the neutron star switched
on as a radio pulsar. Apart from the rapid rotation, recycled pulsars have
weak magnetic fields and are often acccompanied by an undermassive
($\ltap0.4 M_{\odot}$) white dwarf, and sometimes by an ordinary
white dwarf, the remnant of the donor
star. Their X-ray emission is analyzed by Becker \&\ Tr\"umper (1999).

In the galactic disk, all binary X-ray sources,
including recycled pulsars, evolve from primordial binaries (e.g. Verbunt 
1993).
In globular clusters, two additional processes are possible:
tidal capture and exchange encounter (reviewed by Hut et al.\ 1992).
In a tidal capture, one star (in particular a neutron star
or white dwarf) raises tides on a main-sequence or
(sub)giant star in a close
passage; the energy of the tides is taken from the relative motion,
and the stars become bound, as the tidal energy is dissipated.
In an exchange encounter, a star encounters a binary and forms a 
temporary triple with it; one star is then expelled (usually the
star with the lowest mass) and the other two remain bound. In this
way, a neutron star or white dwarf can be exchanged into a binary.
The frequency of occurrence of both tidal capture and exchange
encounters, the {\em collision number},
scales roughly with the square of the number
density of stars and with the volume of the core, where most
encounters occur. 

For the formation of X-ray sources in globular clusters
the relative importance of evolution of primordial binaries on one hand, and
stellar encounters on the other hand,  depends on the source type,
and on the encounter frequency in the cluster.
Many primordial binaries evolve into RS CVn systems; therefore 
all RS CVns in globular clusters
are probably also formed from the evolution of a primordial binary.
Extremely few primordial binaries evolve into low-mass X-ray binaries
(including soft X-ray transients) and on to binaries with recycled
pulsars; therefore all X-ray sources in globular clusters 
with a neutron star are likely formed from stellar encounters.
The situation with cataclysmic variables is intermediate between these
extremes. In a low-density cluster with a high total number of stars,
like $\omega$~Cen, most cataclysmic variables may have evolved
from primordial binaries.
In a cluster with a dense core where stellar encounters are frequent,
like 47~Tuc, most cataclysmic variables may have formed
in stellar encounters.

\begin{figure}[]
\centerline{
\parbox[b]{8.0cm}{\psfig{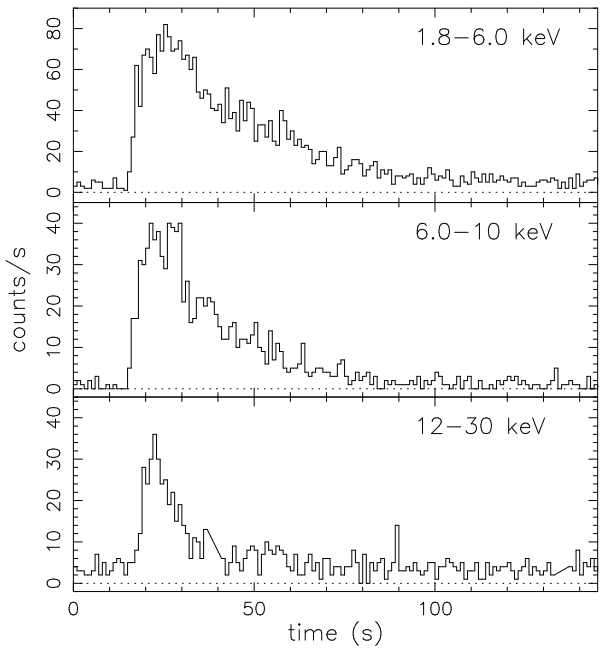}}
\parbox[b]{6.0cm}{\caption{X-ray burst of NGC~6440 observed on
26 Aug 1998 with BeppoSAX. The faster decline at higher energies
is evidence of a thermonuclear burst, and shows that the X-ray
source is a neutron star. Adapted from in 't Zand et al.\ (1999).
\label{fvfigf}}}}
\end{figure}

The situation in reality is more complicated and less well understood
than the above description suggests, as discussed in more detail
by Davies in this volume. The efficiency of tidal capture
and the number and properties of binaries suitable for exchange
encounters in globular clusters are uncertain. The number of
neutron stars in globular clusters is uncertain by an order of 
magnitude. A binary formed by a stellar
encounter in a very dense cluster can be destroyed again in a subsequent
encounter. 
In addition, many aspects of binary evolution are not well understood:
estimates for the life time of low-mass X-ray binaries or of cataclysmic
variables differ by an order of magnitude.
It must be concluded that the apparent accuracy of numbers produced in 
detailed simulations of the formation of X-ray binaries in globular 
clusters is spurious.

\section{Luminous sources}
\begin{figure}[]
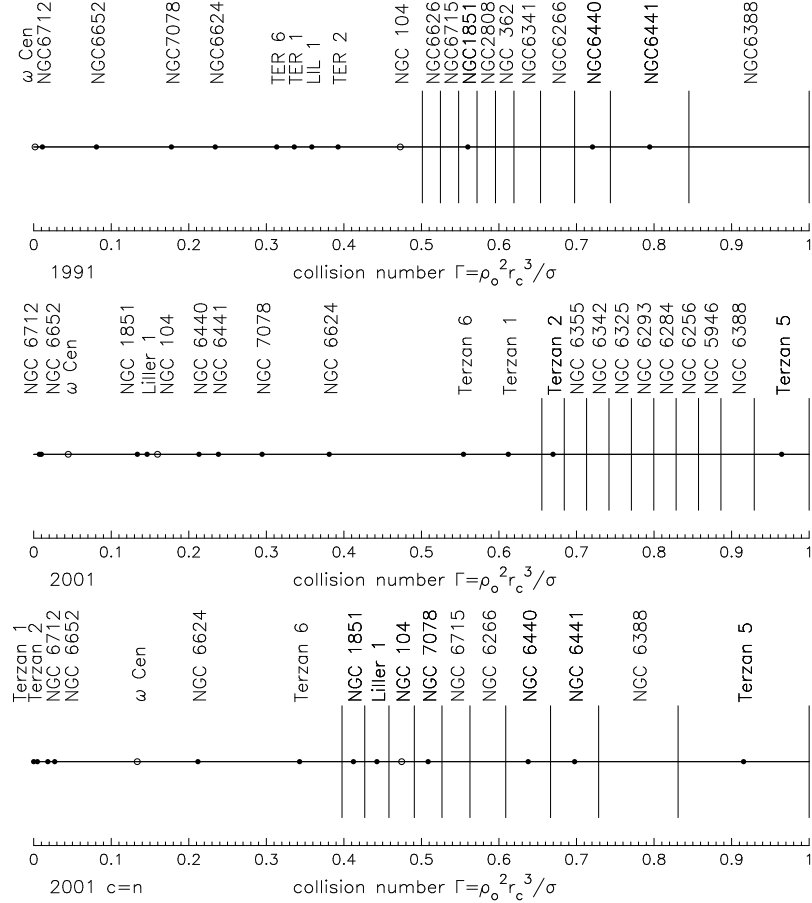

\centerline{\psfig{figure=verbuntfig3a.ps,width=0.8\columnwidth,clip=t}}
\centerline{\psfig{figure=verbuntfig3b.ps,width=0.8\columnwidth,clip=t}}
\centerline{\psfig{figure=verbuntfig3c.ps,width=0.8\columnwidth,clip=t}}
\caption{To test the collision hypothesis for bright X-ray sources, one
orders the globular clusters on their collision number $\Gamma$,
and assigns them a line section proportional to $\Gamma$. The clusters with an
X-ray source should then be distributed homogeneously along the line.
The top figure reproduces this test from Predehl et al.\ (1991), based on
cluster parameters from Chernoff \&\ Djorgovski (1989; their list does not
include Terzan~5);
a $\bullet$ indicates a cluster with a bright X-ray source; for comparison
$\omega$~Cen and 47~Tuc are indicated $\circ$. The ten clusters with largest
$\Gamma$ are delineated.
The middle graph is an update for new cluster parameters (Harris 1996;
version of June 22, 1999). In both, collapsed clusters are
treated approximately by setting the central density
$\rho_{\circ}$ to $10^6M_{\odot}$\,pc$^{-3}$ and the core radius $r_c$ to
0.1\,pc, i.e.\ all collapsed clusters have the same collision number.
In the lower graph, the collision number for collapsed clusters is computed
from their central density and core radius as for other clusters, to
illustrate the effect of different treatments of collapsed clusters.
All three versions pass the Kolmogorov-Smirnov test for a homogeneous
distribution.
\label{fvfigb}}
\end{figure}

A luminous X-ray binary may have a neutron star or a black hole as the
accreting star. In the course of detailed studies of the cluster
sources in our galaxy, it has been found that 11 of the 12
(now: 11 of 13; see below) luminous cluster sources show X-ray
bursts (the most recent addition is NGC~6440, see Figure\,\ref{fvfigf}). 
Such bursts are interpreted as sudden thermonuclear fusion 
into carbon of a layer of helium on the surface of a neutron star; thus
every burster is a neutron star.
The dearth of black hole accretors in globular clusters has been
explained as follows. Once the more massive stars in a cluster have
evolved, black holes are the most massive objects, sink towards
the cluster center, and turn into black-hole binaries via exchange
encounters. Encounters between black-hole binaries lead to large
recoil velocities and virtually all black holes are thus shot out
of the cluster (Portegies Zwart \& McMillan 2000).

\begin{figure}[]
\centerline{
\parbox[b]{8.0cm}{\psfig{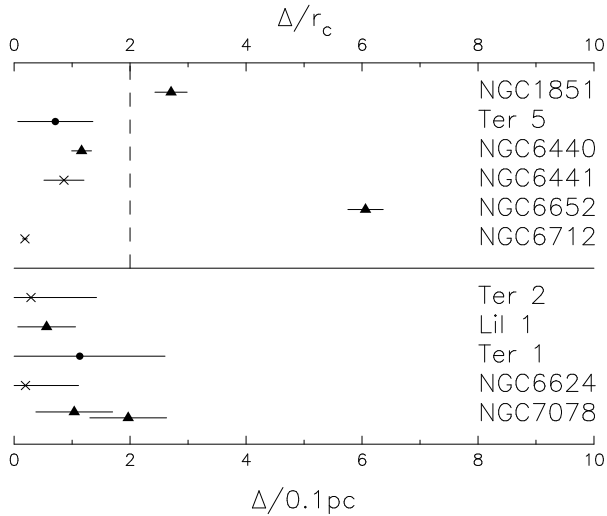}}
\parbox[b]{6.0cm}{\caption{Distance to the cluster center of the bright
X-ray sources, in units of the core radius; or of 0.1\,pc for collapsed
clusters. $\times$, $\bullet$, and $\triangle$ indicate Einstein, Rosat,
and Chandra positions (the latter taken from Homer et al.\ 2001a,b,
Heinke et al.\ 2001, Pooley et al.\ 2001b, White \&\ Angelini 2001). 
Uncertainties include an assumed $1\farcs2$ uncertainty in the cluster center.
\label{fvfigh}}}}
\end{figure}

If the formation of low-mass X-ray binaries is due to close stellar encounters,
one predicts that the probability of a cluster to contain such a binary
is proportional to its collision number
$\Gamma\equiv{\rho_{\circ}}^2{r_c}^3/\sigma$, where
$\rho_{\circ}$ and $\sigma$ are the central density and velocity dispersion, 
and $r_c$ is the core radius.
One also predicts that most such sources
reside in or near the cluster core. It appears that these predictions
are correct (Verbunt \&\ Hut 1987), as illustrated in Figures\,\ref{fvfigb},
\ref{fvfigh}.
It should be noted that precise tests are difficult, given the uncertainty
in how to treat collapsed clusters; and in the values of the cluster 
parameters. Three versions of the test are illustrated in Figure\,\ref{fvfigb}.
The effect of the uncertainty in cluster parameters is illustrated by the
difference between the top two graphs, which use the same equations, but
different cluster catalogues. The effect of different treatments of
collapsed clusters is illustrated by the difference between the lower two
graphs. In the lower graph, collapsed clusters are treated like ordinary
clusters; the presence of bright sources in Terzan 1 and 2 is then unexpected,
as these clusters have a low collision number. In the top two graphs,
the central density and core radius, and thus the collision number,
of any collapsed cluster is set to a
fixed number -- a rather arbitrary procedure.
Alternatively (and better justified) one can apply the test to ordinary,
non-collapsed clusters only (see Predehl et al.\ 1991).

The dominant variation of the collision numbers is due to the variation
in central density over more than 6 orders of magnitude between clusters;
in contrast, the core radii (expressed in parsecs) and velocity dispersions
vary much less (by $\sim 2$ and $\ltap1$ order of magnitude respectively).
It may be noted also that the velocity dispersion is not taken from
observations, but computed assuming a King model as 
$\sigma\propto r_c\sqrt{\rho_{\circ}}$; effectively therefore
$\Gamma\propto\rho_o^{1.5}r_c^2$.


Of the five orbital periods now known (most recent additions: NGC~6440,
in 't Zand et al.\ 2000, and NGC\,6712, Homer et al.\ 1996) two are
ultra-short at 11.4 and 20.6 minutes, which indicates white dwarf donors.
Two other systems may have such short periods as well (Homer et al.\ 2001a).
This high fraction requires explanation.
Comparison of NTT observations made before and during the 1998 outburst of the 
transient bright X-ray source in NGC~6440 provided an optical counterpart,
the first one found for a transient in a cluster (Verbunt et al.\ 2000).

Chandra has already produced two interesting results for the bright
cluster sources in our galaxy. First, White \&\ Angelini (2001)
have found that M~15 contains two bright sources, so close that
earlier satellites couldn't separate them. This resolves the
puzzling combination of properties found for the M~15 source:
on the one hand, the extended partial eclipse of the X-ray source
indicated that the central source is not directly observed and that
the X-rays observed on earth are scattered from a corona above the
accretion disk; on the other hand a thermonuclear burst of the source
reached the Eddington limit, proving that the neutron star is
observed directly. Apparently, one source is the disk corona source,
and the other source is the burster.
The occurrence of two sources was expected in 
clusters with very high collision numbers rather
than in M~15 (NGC\,7078) with its not very high collision number
(see Fig.\,\ref{fvfigb}).
The occurrence of two sources in M~15 thus suggests a higher number
of potential bright X-ray sources in globular clusters than hitherto
suspected.
The absence of a bright source in $\omega$~Cen is not surprising,
given its low collision number.

The second interesting Chandra result is its proof, thanks to a more
accurate position, that the bright source in NGC~6652 is at 6 core radii from
the cluster center (Heinke et al.\ 2001), which indicates that this source
is the result of an exchange encounter rather than of a tidal capture.

The improved sensitivity and high spatial resolution of Chandra
has led to the detection of many X-ray sources in the globular clusters
of other galaxies, including NGC~4697 (Sarazin et al.\ 2000) and
NGC~1399 (Angelini et al.\ 2001). Remarkably, some clusters are very
bright, $\gtap 10^{39}$ erg/s. This, and the softness of their
X-ray spectra, suggests that these cluster sources are black holes.
Clusters with luminosities $\gtap 10^{38}$ erg/s may be examples
of clusters with several bright low-mass X-ray binaries.
The occurrence of such bright sources in clusters of these
galaxies is probably a consequence of the very large numbers
of clusters that these galaxies have, i.e.\ of the presence of
clusters with more extreme properties than the clusters in our
galaxy (in terms of mass, central density and core size). 
The luminosity distribution of bright globular cluster sources
in our galaxy is dominated by small-number statistics, which hampers
statistical comparison with the bright sources in clusters of other galaxies.

\section{Dim sources}

Dim sources in globular clusters were first detected with Einstein
by Hertz \&\ Grindlay (1983), who suggested that most of them are cataclysmic
variables. From the observation that cataclysmic variables 
in the galactic disk usually have X-ray luminosities $\ltap 10^{33}$
erg/s, Verbunt et al. (1984) suggested that the more luminous dim
sources in globular clusters are neutron stars accreting at a low rate,
i.e.\ quiescent transients. 

\begin{figure}[]
\centerline{\psfig{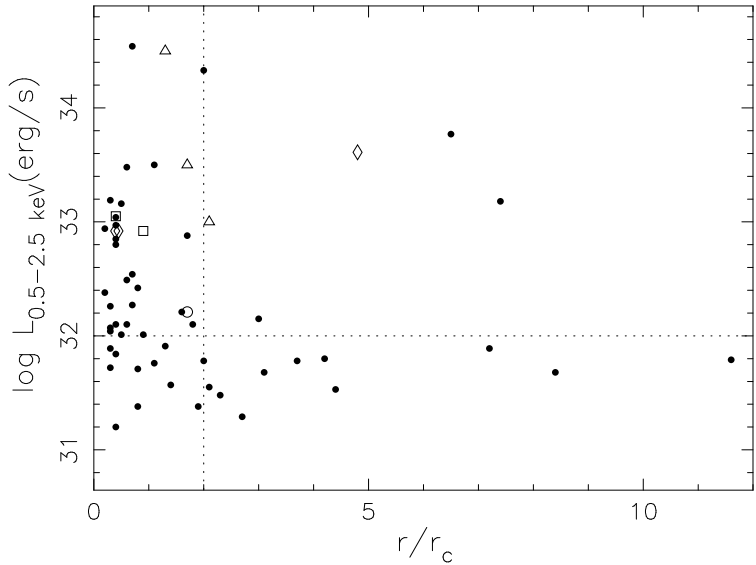}}
\caption{X-ray luminosity of dim sources in globular 
clusters as a function of distance to the cluster center, in units of the
core radius. $\bullet$ ROSAT source, $\Box,\circ,$ sQXT in 47~Tuc, 
$\omega$~Cen; $\triangle,\diamondsuit$ Chandra sources in 
Liller~1 and NGC~6652. The horizontal dotted line
indicates the approximate limit to the X-ray luminosity of cataclysmic 
variables in the galactic disk. A source at $r=26r_c$ is omitted.
\label{fvfigc}}
\end{figure}

A complete census of all 57 dim sources detected with ROSAT is provided
by Verbunt (2001). Figure\,\ref{fvfigc} shows the X-ray luminosity of
the sources from that census, as a function of distance to the cluster
center. Most sources are within 2 core radii from the center.  A
significant population of dim sources is also present further out,
especially at luminosities $\ltap10^{32}$ erg/s; some of these sources
may be primordial cataclysmic variables and RS CVn systems.  At
luminosities $\gtap10^{33}$ erg/s, where most sources are probably
quiescent transients, two ROSAT sources are detected well outside the
core: these may have formed via exchange encounters.  The ROSAT data
indicated that the X-ray spectra of globular clusters are soft,
compatible with those of quiescent transients.

One remarkable result from the ROSAT census is that the X-ray emission
of most globular clusters per unit mass is lower than that of the old open
cluster M~67 (Verbunt 2001). Our understanding of this result is
not helped by the fact that the three most luminous X-ray sources in
M~67 are mysterious: one is a triple containing two blue stragglers;
two are located below the subgiant branch
in the Hertzsprung Russel diagram and we do not understand their
evolutionary status (Belloni et al.\ 1998, van den Berg et al.\ 2001).

\begin{figure}[]
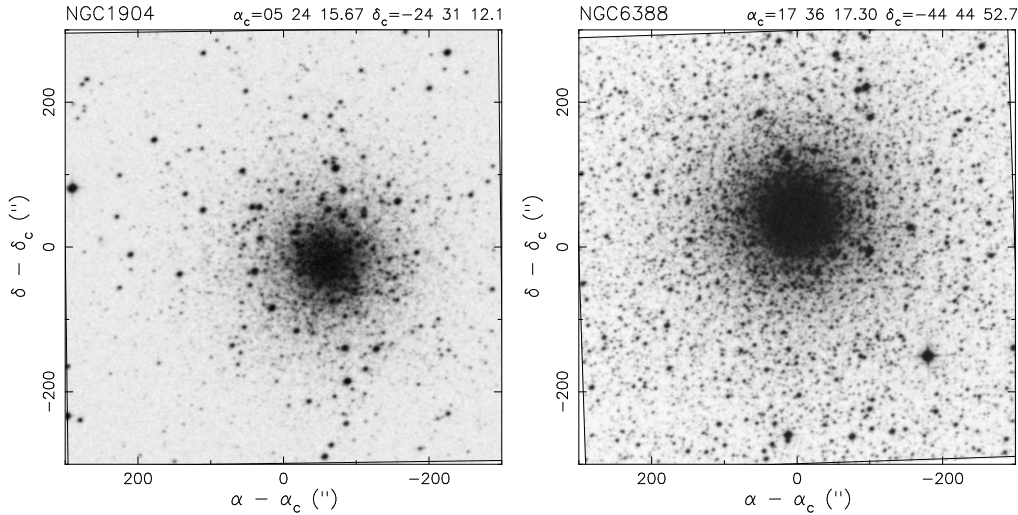

\centerline{
\parbox[b]{6.7cm}{\psfig{figure=verbuntfig6a.ps,width=6.7cm,clip=t}}
\parbox[b]{6.7cm}{\psfig{figure=verbuntfig6b.ps,width=6.7cm,clip=t}}}
\caption{Positions of two dim sources, with $L_{\rm x}\gtap10^{33}$ erg/s,
discovered with ROSAT in the globular clusters NGC~1904 and
NGC~6388. The images from the Digitized Sky Survey are centered on the
X-ray positions, to illustrate the offset to the cluster center.
\label{fvfigg}}
\end{figure}

The spatial resolution of Chandra made possible the first detection
of dim sources in clusters with bright sources, viz.\ in Liller~1
and in NGC~6652 (Homer et al.\ 2001b, Heinke et al.\ 2001).
The luminosities of several of these suggests that they are quiescent
transients, one well outside the core (see Figure\,\ref{fvfigc}).

The superior spatial resolution of Chandra further resulted in the 
discovery of dozens of sources in each of five clusters devoid of bright 
sources.

\begin{figure}[]
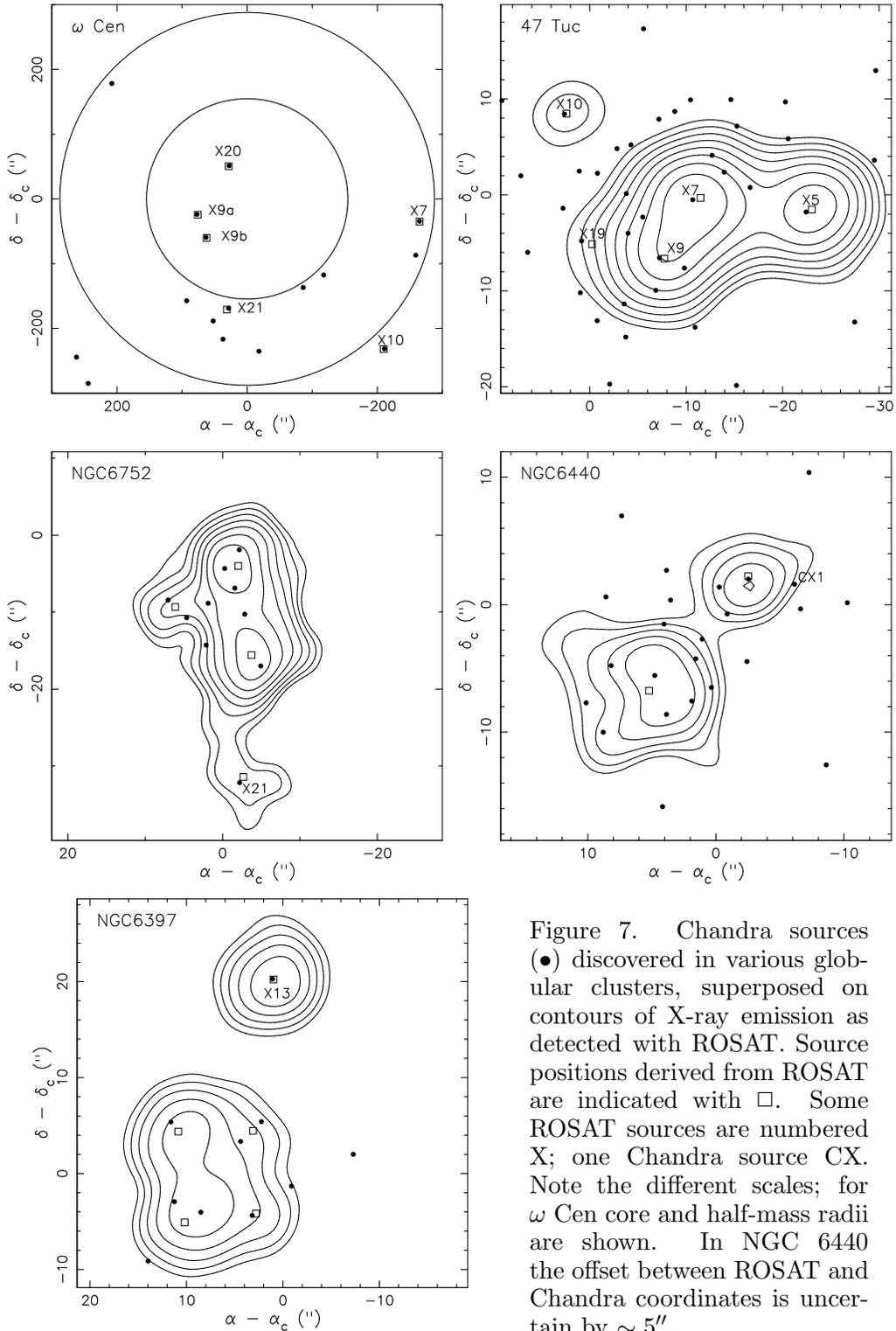

\centerline{
\parbox[b]{6.7cm}{\psfig{figure=verbuntfig7a.ps,width=6.7cm,clip=t}}
\parbox[b]{6.7cm}{\psfig{figure=verbuntfig7b.ps,width=6.7cm,clip=t}}}
\centerline{
\parbox[b]{6.7cm}{\psfig{figure=verbuntfig7c.ps,width=6.7cm,clip=t}}
\parbox[b]{6.7cm}{\psfig{figure=verbuntfig7d.ps,width=6.7cm,clip=t}}}
\centerline{
\parbox[b]{6.7cm}{\psfig{figure=verbuntfig7e.ps,width=6.7cm,clip=t}}
\parbox[b]{6.7cm}{
\caption{Chandra sources ($\bullet$) discovered in various globular clusters,
superposed on contours of X-ray emission as detected with ROSAT.
Source positions derived from ROSAT are indicated with $\Box$.
Some ROSAT sources are numbered X; one Chandra source CX. Note the different
scales; for $\omega$~Cen core and half-mass radii are shown. In NGC~6440
the offset between ROSAT and Chandra coordinates is uncertain by $\sim5''$.
\label{fvfige}}
}}
\end{figure}

In $\omega$~Cen, a low-density cluster with a large core, ROSAT was already
able to separate the X-ray sources, as shown in Figure\,\ref{fvfige};
Chandra has added new sources to bring the total to 40 (Rutledge
et al.\ 2001) or more (Cool, these proceedings). The brightest X-ray source in
$\omega$~Cen (X7 from Verbunt \&\ Johnston 2000, see Fig.\,\ref{fvfige}) 
has a relatively soft
spectrum, and probably is a quiescent soft X-ray transient.  
The presence of such a source
in $\omega$~Cen is surprising: as argued in Sect.\,2 above, binaries
with neutron stars in globular clusters are most likely formed in
close stellar encounters, but the number of such encounters in
$\omega$~Cen is relatively low ($\sim$0.2\%\ of the encounter rate in
all clusters together).  Unless the
soft X-ray transient in $\omega$~Cen is a statistical fluke, its
existence indicates a large number of quiescent transients in
the globular cluster system. Alternatively, the system could have evolved
from a primordial binary ($\omega$~Cen has a mass of $\sim3\,10^6M_{\odot}$,
see Meylan, these proceedings). This would indicate a rather
larger fraction of primordial binaries evolving into binaries with a
neutron star than hitherto thought possible; and thus also implies a large
number of quiescent X-ray transients in the galactic disk.  

In higher-density clusters, ROSAT failed to resolve all X-ray sources,
even with sophisticated software, as illustrated in Figure\,\ref{fvfige}:
ROSAT did detect the brighter sources -- and with these, the bulk of
the X-ray flux -- but in their glare could not detect
the fainter sources. The first Chandra results have found many, many
dim sources: more than 100 in 47~Tuc alone (Grindlay et al.\ 2001a);
19 in NGC~6752 (Pooley et al.\ 2001a), 25 in
NGC~6397 (Cool, these proceedings, Grindlay et al.\ 2001b\footnote{The
latter reference was published after the meeting; I include its results,
also in Fig.\,\ref{fvfige}, for completeness}),
and 24 in NGC~6440 (Pooley et al.\ 
2001b). Most of these sources are cluster members, rather than
fore- or background objects. Of them, 2 in 47~Tuc (X5, X7 in 
Fig.\,\ref{fvfige}), 1 in NGC\,6397 (X13 in Fig.\,\ref{fvfige}) and
some 5 in NGC~6440 are quiescent soft X-ray transients. (Perhaps it is safer
to say that they are neutron stars accreting at low luminosity, since
we do not know when or whether they ever show outbursts.\footnote{In fact,
during the conference one of the sources in NGC~6440, CX1 in 
Fig.\,\ref{fvfige}, did go into outburst! see in 't Zand et al.\ (2001)})

\begin{figure}[]
\centerline{\psfig{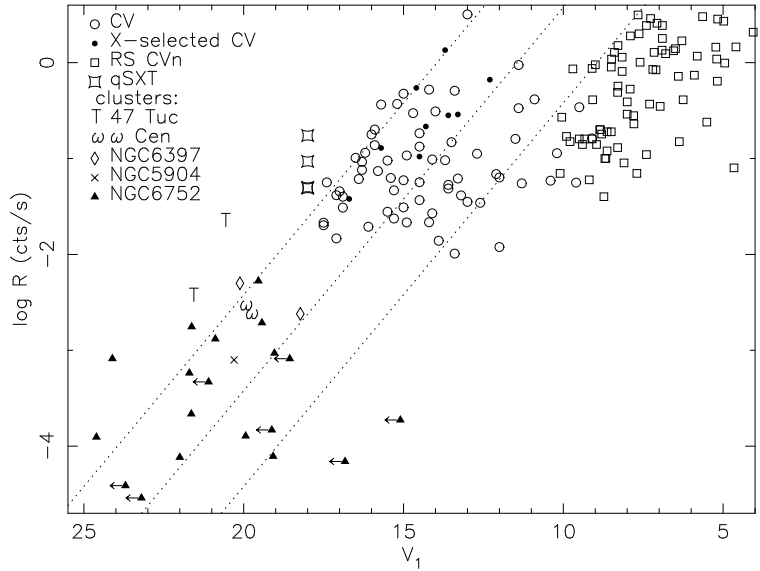}}
\caption{
X-ray flux, expressed in ROSAT PSPC countrate $R$, as a function of
visual magnitude, for cataclysmic variables, RS CVns and quiescent
soft X-ray transients in the galactic disk, and for various
optically identified dim X-ray sources in globular clusters.
For NGC~6752 $R$ is computed from Chandra countrates; and arrows
indicate upper limits to the visual flux (from Pooley et al.\ 2001a).
The dotted lines indicate constant ratios of X-ray to optical flux.
\label{fvfigd}}
\end{figure}

Most ROSAT positions have accuracies of $\sim 5''$,
which in a dense cluster core is insufficient for unambiguous
optical identifications, as shown by wrong identifications
in e.g.\ 47~Tuc and M~13 (discussed on the basis of more accurate ROSAT
positions in Verbunt \&\ Hasinger 1998,
Verbunt 2001). However, objects with H\,$\alpha$ emission
and variability on time scales of hours -- presumed to be cataclysmic
variables -- have been suggested as counterparts for other ROSAT sources in 
47~Tuc and NGC~6397, and were confirmed with Chandra
(Grindlay et al.\ 2001a,b). Two variables in NGC~6752 have been
identified with X-ray sources, after correction of their optical
positions (Pooley et al.\ 2001a).
The optical to X-ray flux ratio of dim sources can be compared with that of
the various proposed types of counterparts; this is done in 
Figure\,\ref{fvfigd}. The X-ray to
optical flux ratio of optically identified sources in $\omega$~Cen
(Carson et al.\  2000), NGC~5904 (Hakala et al.\ 1997) and NGC~6397
(Cool et al.\ 1995, Verbunt \&\ Johnston 2000)
are in the range found for cataclysmic variables in the disk.
These sources thus probably are cataclysmic variables. The source in
NGC~5904 is interesting in being located in the outskirts of the cluster,
at 11.6 core radii; it is a dwarf nova. Two sources
in 47~Tuc (X9 and X19 of Verbunt \&\ Hasinger 1998, see Fig.\,\ref{fvfige}) 
and one in NGC~6752 (CX6 of Pooley et al.\ 2001a, X21 in Fig.\,\ref{fvfige}) 
have an X-ray to optical flux ratio higher than
those for cataclysmic variables in the disk, and more similar to those of
the quiescent soft X-ray transients Cen X-4 and Aql X-1. However, the
Chandra spectra of these sources are harder than the
quiescent X-ray spectra of transients. Thus these sources are probably
cataclysmic variables.
The presence of H\,$\alpha$ emission lines is also used as an indicator
that a source is a cataclysmic variable. Whereas this conclusion is often
correct, it is worth noting that quiescent X-ray transients also may show
substantial H\,$\alpha$ emission lines (for example Cen X-4, van Paradijs
et al.\ 1987). H\,$\alpha$ emission by itself is not definite proof
for a cataclysmic variable.

\section{Conclusions}

The distribution of bright X-ray sources over clusters with different
encounter rates, as expressed in the collision number, is compatible
with the hypothesis that these sources are formed in close encounters
of neutron stars with other cluster stars or with binaries.
The spatial distribution of low-luminosity sources in globular
clusters shows a concentration towards the core, again compatible
with the hypothesis that many sources are formed in close encounters.
Sources far away from the cluster centers may have evolved from primordial
binaries, or -- if they contain a neutron star -- from a close
encounter of a neutron star with a binary, the recoil of which
brought them to their current location.
For future studies, a calculation of encounter rates in collapsed clusters
would be very useful; and more generally calculations of encounter rates
in evolving clusters. After all, the current encounter rate in a cluster
need not be representative for sources formed long ago.

$\omega$~Cen was not expected to contain binaries with a neutron
star: its encounter rate is relatively low, and in agreement with
this, no bright X-ray source or recycled radio pulsars have been
found in $\omega$~Cen. The discovery of a binary in which -- based on
its luminosity and soft X-ray spectrum --  a neutron star accretes
at a low-rate from a companion suggests that there may be rather
more binaries with neutron stars in the galactic globular cluster
system than previously expected.
The major uncertainties in theoretically predicting the formation
rates of such binaries are the uncertainty in the number of
neutron stars formed with velocities sufficiently small that they 
are retained in the globular cluster; and the cross sections of the
various mechanisms which may put a neutron star in a binary.

A new era has started with Chandra. The study of large numbers of
globular clusters in other galaxies will hopefully elucidate the
relation between cluster properties -- total mass, central density,
and core radius -- and X-ray properties. In particular, the discovery of
accreting black holes in these clusters is interesting.
The accurate positions of sources in clusters in our own galaxy enables
unambigious identifications with radio and optical sources.
A first result from this is that several of the cataclysmic
variables in globular clusters have a higher X-ray to optical flux ratio
than cataclysmic variables in the galactic disk.

\end{document}